\documentclass{iaus}
\usepackage{graphicx}
\usepackage{amsmath}
\usepackage{amssymb}
\usepackage{amstext}
\usepackage{graphicx}
\usepackage{epstopdf}
\usepackage{epsfig}
\usepackage{color}
\definecolor{myColor}{rgb}{0.9,0.9,0.9}    
\newcommand{\apj}{\textit{ApJ}}

\newcommand{\aj}{\textit{AJ}}

\begin{document}

\title{A Search for Exotrojans in Transiting Exoplanetary systems}
\author{N. Madhusudhan \& Joshua N. Winn}

\affiliation{Department of Physics and Kavli Institute for 
Astrophysics and Space Research, MIT, Cambridge, MA 02139; {\tt 
nmadhu@mit.edu}}

\maketitle

\begin{abstract}
  We present a search for Trojan companions to 25
  transiting exoplanets. We use the technique of Ford \& Gaudi, in
  which a difference is sought between the observed transit time and the
  transit time that is calculated by fitting a two-body Keplerian
  orbit to the radial-velocity data. This technique is sensitive to
  the imbalance of mass at the L4/L5 points of the planet-star
  orbit. No companions were detected. The median 2$\sigma$ upper limit
  is 60~$M_\oplus$, and the most constraining limit is 2.5 $M_\oplus$
  for the case of GJ~436. \end{abstract}

\keywords{celestial mechanics --- techniques: photometric --- techniques: radial velocities}

\firstsection

\section{Introduction}

Trojan companions are bodies in a 1:1 mean-motion resonance with a planet, 
librating around one of the two Lagrange points (L4 and L5) of the planet's orbit 
around the star. Several methods have been proposed to detect Trojan companions to
exoplanets (Laughlin \& Chambers 2002, Croll et al.~2007, Ford \&
Holman 2007,  Ford \& Gaudi 2006). In this paper, we present a search for 
Trojan companions to 25 known transiting exoplanets for which 
suitable data are available, using the method of Ford \& Gaudi (2006, hereafter,
``FG''). An important virtue of this method is that a sensitive search for Trojans can be 
performed using only the RV and photometric data that are routinely 
obtained while confirming transit candidates and characterizing the planets. 

\section{Method}
\label{sec:method}

The basic idea of the FG method is to compare the photometrically 
observed transit time ($t_O$) with the expected transit time ($t_C$) that is 
calculated by fitting a two-body Keplerian orbit to the RV data, i.e. 
assuming no Trojan. The presence of a Trojan companion as a third body
would cause a timing offset $\Delta t = t_O - t_C$.

In the case of a circular orbit, if there is no Trojan companion, the force
vector on the star points directly at the planet and the observed
transit time $t_O$ coincides with the time $t_V$ when the orbital
velocity of the star is in the plane of the sky (i.e., the time of RV null). 
If instead there is a single Trojan located at the L4 or L5 Lagrange point 
(or librating with a small amplitude), then the force vector on the star does not
point directly at the planet; it is displaced in angle toward the
Trojan companion. As a result, $t_O$ occurs earlier or later than
$t_V$. The magnitude of $t_O - t_V$ is proportional to the Trojan mass
$m_T$, for small values of the Trojan-to-planet mass ratio (Ford \&
Gaudi 2006):
$\Delta t \simeq \pm 37.5~\text{min}~(P/3 \, \textrm{days}) (m_T/10 \, M_\oplus)(0.5\,M_J/(m_P + m_T))~~(1)$.
The positive sign corresponds to a mass excess at the L4 point
(leading the planet) while the negative sign corresponds to that  
at the L5 point (lagging the planet). For an eccentric two-body orbit, $t_C$ does 
not generally coincide with $t_V$. We calculate $t_C$ by fitting a two-body 
Keplerian orbit to the RV data and calculating the expected transit time based 
on the fitted orbital parameters. 

\section{Data Analysis}
\label{sec:results}

The RV data were taken from the available literature on each
system. We fitted the Keplerian model to the RV data using a Markov Chain Monte
Carlo (MCMC) technique, employing a Metropolis-Hastings algorithm
(see, e.g., Ford~2005). In general, uniform priors were 
used for all parameters. A prior constraint on $e\cos\omega$ was used 
where secondary eclipse measurements are available in literature. 
A single chain of $\sim 10^6$ links was used for each system. For each parameter, 
we found the mode of the {\it a posteriori}\, distribution, and the 68.3\% confidence 
interval, defined as the range that excludes 15.9\% of the probability at each extreme of 
the distribution. We thus obtained $t_C$ from fitting the RV data. We determined $t_O$ using the most precise  published photometric ephemeris, and then computed $\Delta t = t_O - t_C$. The results for $\Delta t$ are translated into constraints on 
the Trojan mass $m_T$ using Eq.~(1) for circular orbits, and using equivalent expressions 
for eccentric orbits obtained by numerical integrations of three-body orbits. 

\section{Results}
\label{sec:results}

The 95.4\% (2$\sigma$) upper limits on $m_T$ and on $m_T/m_P$ are shown in
Figure~1. The median upper limit on $m_T$ is 60~$M_\oplus$, with the most
constraining limit of 2.5~$M_\oplus$ holding for GJ~436. This powerful upper 
limit is possible for this case because of the small stellar and planetary masses, 
and the copious RV data available for this system. The median upper limit on the
mass ratio $m_T/m_P$ is 0.1, with a best-case value of 0.015 for
HD~17156. The powerful constraint in this case is a result of the
unusually long orbital period of 21~days and plentiful precise RV data.

\begin{figure}[h]
\begin{center}
\includegraphics[width=\textwidth]{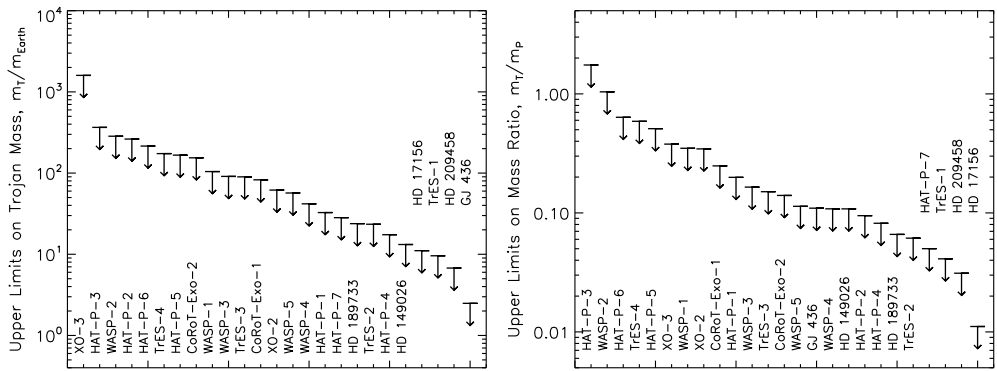}
\caption{95.4 \% upper-limits on trojans masses and on trojan-planet mass ratios. 
The systems are ordered from least constrained to most well constrained, going from left 
to right of the figure.}
\label{fig:trojans}
\end{center}
\end{figure}

In all cases but two, the result for $\Delta t$ was consistent with zero within
2$\sigma$. The exceptions were CoRoT-Exo-2 and WASP-2, for which $\Delta t 
= 30^{+17}_{-14}$ and $-142^{+53}_{-44}$~minutes respectively. 
The results for these two systems are worth following up with additional 
RV data. However, in a sample of 25 systems, even if $\Delta t$ is always consistent with 
zero, one expects approximately one 2$\sigma$ outlier. There were no cases in 
which $\Delta t$ was inconsistent with zero at the 3$\sigma$ level. Hence, we 
conclude that there is no compelling evidence for a Trojan companion in this ensemble.

\end{document}